\documentclass[11pt]{article}
\usepackage{amsmath, amssymb, amsfonts, amsthm}

\newtheorem{theorem}{Theorem}[section]
\newtheorem{proposition}[theorem]{Proposition}
\newtheorem{corollary}[theorem]{Corollary}
\newtheorem{lemma}[theorem]{Lemma}

\newcommand{\rd}{{\rm d}}

\newcommand{\bx}{{\bf x}}

\newcommand{\bR}{{\mathbb R}}

\newcommand{\tr}{\mbox{Tr}}
\newcommand{\wt}{\widetilde}

\newcommand{\cB}{{\cal B}}
\newcommand{\cH}{{\cal H}}

\begin{document}

\title{Derivation of the Gross-Pitaevskii Hierarchy}
\author{Benjamin Schlein\thanks{E-mail: schlein@mathstanford.edu.
Supported by NSF Postdoctoral Fellowship.}\\
\\
Department of Mathematics, Stanford University\\ Stanford, CA
94305, USA}

\maketitle

\begin{abstract}
We report on some recent results regarding the dynamical behavior
of a trapped Bose-Einstein condensate, in the limit of a large
number of particles. These results were obtained in \cite{ESY}, a
joint work with L. Erd\H os and H.-T. Yau.
\end{abstract}

\section{Introduction}

In the last years, progress in the experimental techniques has
made the study of dilute Bose gas near the ground state a hot
topic in physics. For the first time, the existence of
Bose-Einstein condensation for trapped gases at very low
temperatures has been verified experimentally. The experiments
were conducted observing the dynamics of Bose systems, trapped by
strong magnetic field and cooled down at very low temperatures,
when the confining traps are switched off. It seems therefore
important to have a good theoretical description of the dynamics
of the condensate. Already in 1961 Gross \cite{G1,G2} and
Pitaevskii \cite{P} proposed to model the many body effects in a
trapped dilute Bose gas by a nonlinear on-site self interaction of
a complex order parameter (the condensate wave function $u_t$).
They derived the Gross-Pitaevskii equation
\begin{equation}\label{eq:GPE1}
i\partial_t u_t = -\Delta u_t + 8\pi a_0 |u_t|^2 u_t
\end{equation}
for the evolution of $u_t$. Here $a_0$ is the scattering length of
the pair interaction. A mathematically rigorous justification of
this equation is still missing. The aim of this article is to
report on recent partial results towards the derivation of
(\ref{eq:GPE1}) starting from the microscopic quantum dynamics in
the limit of a large number of particles. Here we only expose the
main ideas: for more details, and for all the proofs, we refer to
\cite{ESY}.

Also in the mathematical analysis of dilute bosonic systems some
important progress has recently been made. In \cite{LY1}, Lieb and
Yngvason give a rigorous proof of a formula for the leading order
contribution to the ground state energy of a dilute Bose gas (the
correct upper bound for the energy was already obtained by Dyson
in \cite{Dy}, for the case of hard spheres). This important result
inspired a lot of subsequent works establishing different
properties of the ground state of the Bose system. In \cite{LSY1},
the authors give a proof of the asymptotic exactness of the
Gross-Pitaevskii energy functional for the computation of the
ground state energy of a trapped Bose gas. In \cite{LS}, the
complete condensation of the ground state of a trapped Bose gas is
proven. For a review of recent results on dilute Bose systems we
refer to \cite{LSSY2}. All these works investigate the properties
of the ground state of the system. Here, on the other hand, we are
interested in the dynamical behavior.

Next, we want to describe our main result in some details. To this
end, we need to introduce some notation. From now on we consider a
system of $N$ bosons trapped in a box $\Lambda \subset \bR^3$ with
volume one and we impose periodic boundary conditions. In order to
describe the interaction among the bosons, we choose a positive,
smooth, spherical symmetric potential $V (x)$ with compact support
and with scattering length $a_0$.

Let us briefly recall the definition of the scattering length
$a_0$ of the potential $V(x)$. To define $a_0$ we consider the
radial symmetric solution $f(x)$ of the zero energy one-particle
Schr\"odinger equation
\begin{equation}\label{eq:scatt}
\left( - \Delta + \frac{1}{2} V(x) \right) f(x) = 0,
\end{equation} with the condition $f(x) \to 1$ for $|x| \to
\infty$. Since the potential has compact support, we can define
the scattering length $a_0$ associated to $V(x)$ by the equation
$f(x) = 1 - a_0/|x|$ for $x$ outside the support of $V(x)$ (this
definition can be generalized by $a_0 = \lim_{r \to \infty} r (1 -
f(r))$, if $V$ has unbounded support but still decays sufficiently
fast at infinity). Another equivalent characterization of the
scattering length is given by the formula
\begin{equation}\label{eq:eightpia0}
\int \rd x \, V(x) f(x) = 8 \pi a_0 .
\end{equation}
Physically, $a_0$ is a measure of the effective range of the
potential $V(x)$.

The Hamiltonian of the $N$-boson system is then given by
\begin{equation}\label{eq:ham}
H = - \sum_{j=1}^N \Delta_j + \sum_{i<j} V_a (x_i -x_j)
\end{equation}
with $V_a (x) = (a_0/a)^2 V ( (a_0/a) x )$. By scaling, $V_a$ has
scattering length $a$. In the following we keep $a_0$ fixed (of
order one) and we vary $a$ with $N$, so that when $N$ tends to
infinity $a$ approaches zero. In order for the Gross-Pitaevskii
theory to be relevant we have to take $a$ of order $N^{-1}$ (see
\cite{LSY1} for a discussion of other possible scalings). In the
following we choose therefore $a= a_0 /N$, and thus \( V_a (x) =
N^2 V (Nx) . \) Note that, with this choice of $a$, the
Hamiltonian (\ref{eq:ham}) can be viewed as a special case of the
mean-field Hamiltonian \begin{equation}\label{eq:mf} H_{\text{mf}}
= - \sum_{j=1}^N \Delta_j + \frac{1}{N} \sum_{i<j} \beta^3 V
(\beta (x_i -x_j)) \, .\end{equation} The Gross-Pitaevskii scaling
is recovered when $\beta = N$. We study the dynamics generated by
(\ref{eq:mf}) for other choices of $\beta$ ($\beta = N^{\alpha}$,
with $\alpha < 3/5$) in \cite{EESY}.

Since we have $N$ particles in a box of volume one, the density is
given by $\rho = N$. Hence, the total number of particles
interacting at a given time with a fixed particle in the system is
typically of order \( \rho a^3 \simeq N^{-2} \ll 1 \,.\) This
means that our system is actually a very dilute gas, scaled so
that the total volume remains fixed to one.

The dynamics of the $N$-boson system is determined by the
Schr\"odinger equation
\begin{equation}\label{eq:schr}
i\partial_t \psi_{N,t} = H \psi_{N,t}
\end{equation}
for the wave function $\psi_{N,t} \in L^2 (\bR^{3N}, \rd \bx)$.
Instead of describing the quantum mechanical system through its
wave-function we can describe it by the corresponding density
matrix \(\gamma_{N,t} = |\psi_{N,t} \rangle \langle \psi_{N,t} |\)
which is the orthogonal projection onto $\psi_{N,t}$. We choose
the normalization so that $\tr \, \gamma_{N,t} =1$. The
Schr\"odinger equation (\ref{eq:schr}) takes the form
\begin{equation}\label{eq:hei}
i\partial_t \gamma_{N,t} = [ H , \gamma_{N,t} ]\,.
\end{equation}
For large $N$ this equation becomes very difficult to solve, even
numerically. Therefore, it is desirable to have an easier
description of the dynamics of the system in the limit $N \to
\infty$, assuming we are only interested in its macroscopic
behavior, resulting from averaging over the $N$ particles. In
order to investigate the macroscopic dynamics, we introduce the
marginal distributions associated to the density matrix
$\gamma_{N,t}$. The $k$-particle marginal distribution
$\gamma_{N,t}^{(k)}$ is defined by taking the partial trace over
the last $N-k$ variables. That is, the kernel of
$\gamma_{N,t}^{(k)}$ is given by
\begin{equation*}
\gamma^{(k)}_{N,t} (\bx_k ; \bx'_k) = \int \rd \bx_{N-k} \,
\gamma_{N,t} (\bx_k, \bx_{N,k} ; \bx'_k, \bx_{N-k})
\end{equation*}
where $\gamma_{N,t} (\bx ; \bx')$ denotes the kernel of the
density matrix $\gamma_{N,t}$. Here and in the following we use
the notation $\bx = (x_1 , \dots , x_N)$, $\bx_k = (x_1, \dots ,
x_k)$, $\bx_{N-k} = (x_{k+1} , \dots , x_N)$, and analogously for
the primed variables. By definition, the $k$-particle marginal
distributions satisfy the normalization condition
\begin{equation*}
\tr \, \gamma^{(k)}_{N,t} = 1 \quad \quad \text{for all} \quad
k=1,\dots, N \, .
\end{equation*}
In contrast to the density matrix $\gamma_{N,t}$, one can expect
that, for fixed $k$, the marginal distribution
$\gamma^{(k)}_{N,t}$ has a well defined limit
$\gamma_{\infty,t}^{(k)}$ for $N \to \infty$ (with respect to some
suitable weak topology), whose dynamics can be investigated. In
particular, the Gross-Pitaevskii equation (\ref{eq:GPE1}) is
expected to describe the time evolution of the limit
$\gamma^{(1)}_{\infty,t}$ of the one-particle marginal
distribution, provided $\gamma^{(1)}_{\infty,t} =|u_t \rangle
\langle u_t|$ is a pure state. Eq. (\ref{eq:GPE1}) can be
generalized, for $\gamma_{\infty,t}^{(1)}$ describing a mixed
state, to
\begin{equation}\label{eq:GPE2}
\begin{split}
i\partial_t \gamma^{(1)}_{\infty,t} (x;x') = \; &(-\Delta +
\Delta') \gamma^{(1)}_{\infty,t} (x;x') + 8 \pi a_0 \left(
\gamma_{\infty,t}^{(1)} (x;x) - \gamma_{\infty,t}^{(1)} (x';x')
\right) \gamma^{(1)}_{\infty,t} (x;x') ,
\end{split}
\end{equation}
which we again denote as the Gross-Pitaevskii equation.

To understand the origin of (\ref{eq:GPE2}), we start from the
dynamics of the marginals $\gamma_{N,t}^{(k)}$, for finite $N$.
{F}rom the Schr\"odinger equation (\ref{eq:hei}), we can easily
derive a hierarchy of $N$ equations, commonly called the BBGKY
hierarchy, describing the evolution of the distributions
$\gamma^{(k)}_{N,t}$, for $k=1,\dots,N$:
\begin{equation}\label{eq:BBGKY}
\begin{split}
i\partial_t \gamma_{N,t}^{(k)} (\bx_k ;&\bx'_k) = \sum_{j=1}^k
(-\Delta_{x_j} + \Delta_{x'_j}) \gamma_{N,t}^{(k)} (\bx_k;\bx'_k)
+ \sum_{j \neq \ell}^{k} (V_a (x_j -x_{\ell}) - V_a (x'_j -
x'_{\ell})) \gamma_{N,t}^{(k)} (\bx_k;\bx'_k) \\ &+ (N-k)
\sum_{j=1}^k \int \rd x_{k+1} (V_a (x_j - x_{k+1}) - V_a (x'_j -
x_{k+1}))\, \gamma_{N,t}^{(k+1)} (\bx_k,x_{k+1};\bx'_k,x_{k+1}).
\end{split}
\end{equation}
Here we use the convention that $\gamma^{(k)}_{N,t} =0$, for $k
>N$. Hence, the one-particle marginal density $\gamma^{(1)}_{N,t}$
satisfies
\begin{equation}\label{eq:gamma1}
\begin{split}
i\partial_t \gamma^{(1)}_{N,t} (x_1;x'_1) = \; &(-\Delta_{x_1} +
\Delta_{x'_1}) \gamma_{N,t}^{(1)} (x_1;x'_1) \\ &+ (N-1) \int \rd
x_2 \, \left(V_a (x_1 -x_2) - V_a (x'_1 - x_2) \right)
\gamma^{(2)}_{N,t} (x_1 , x_2; x'_1 , x_2) \, .
\end{split}
\end{equation}
In order to get a closed equation for $\gamma_{N,t}^{(1)}$ we need
to assume some relation between $\gamma_{N,t}^{(1)}$ and
$\gamma_{N,t}^{(2)}$. The most natural assumption consists in
taking the two particle marginal to be the product of two
identical copies of the one particle marginal. Although this kind
of factorization cannot be true for finite $N$, it may hold in the
limit $N \to \infty$. We suppose therefore that
$\gamma_{\infty,t}^{(k)}$, for $k=1,2$, is a limit point of
$\gamma_{N,t}^{(k)}$, with respect to some weak topology, with the
factorization property
\begin{equation*}
\gamma^{(2)}_{\infty,t} (x_1 , x_2 ; x'_1 , x'_2) =
\gamma^{(1)}_{\infty,t} (x_1 ; x'_1) \gamma^{(1)}_{\infty,t} (x_2;
x'_2)\, .
\end{equation*}
Under this assumption we could naively guess that, in the limit $N
\to \infty$, Eq.~(\ref{eq:gamma1}) takes the form
\begin{equation}\label{eq:wrong}
\begin{split}
i\partial_t \gamma_{\infty,t}^{(1)} (x_1; x'_1) = \;
&(-\Delta_{x_1} + \Delta_{x'_1}) \gamma_{\infty,t}^{(1)} (x_1;
x'_1) + \left(Q_t (x_1) - Q_t (x'_1) \right)
\gamma^{(1)}_{\infty,t} (x_1;x'_1)
\end{split}
\end{equation}
with
\begin{equation*}\begin{split}
Q_t (x_1) &= \lim_{N \to \infty} N \int \rd x_2 \, V_a (x_1 - x_2)
\gamma_{\infty,t}^{(1)} (x_2 ; x_2)\\ & = \lim_{N \to \infty} \int
\rd
x_2 \, N^3 V (N(x_1 - x_2)) \gamma_{\infty,t}^{(1)} (x_2 ; x_2)\\
& = b_0 \gamma^{(1)}_{\infty,t} (x_1 ; x_1)
\end{split}
\end{equation*}
where we defined $b_0 = \int \rd x \, V(x)$. Using the last
equation, (\ref{eq:wrong}) can be rewritten as
\begin{equation}\label{eq:GPwrong}
\begin{split}
i\partial_t \gamma^{(1)}_{\infty,t} (x_1;x'_1) = \;
&(-\Delta_{x_1} + \Delta_{x'_1}) \gamma^{(1)}_{\infty,t} (x_1;
x'_1) + b_0 \left( \gamma_{\infty,t}^{(1)} (x_1;x_1) -
\gamma^{(1)}_{\infty,t} (x'_1;x'_1) \right)
\gamma^{(1)}_{\infty,t} (x_1;x'_1)
\end{split}
\end{equation}
which is exactly the Gross-Pitaevskii equation (\ref{eq:GPE2}),
but with the wrong coupling constant in front of the non-linear
term ($b_0$ instead of $8\pi a_0$). The fact that we get the wrong
coupling constant suggests that something was not completely
correct with the naif argument leading from (\ref{eq:gamma1}) to
(\ref{eq:GPwrong}). Reconsidering the argument, the origin of the
error is quite clear: when passing to the limit $N \to \infty$ we
first replaced $\gamma_{N,t}^{(2)}$ with $\gamma_{\infty,t}^{(2)}$
and only after this replacement we took the limit $N \to \infty$
in the potential. This procedure gives the wrong result because
the marginal distribution $\gamma^{(2)}_{N,t}$ has a short scale
structure living on the scale $1/N$, which is the same length
scale characterizing the potential $V_a (x)$. The short scale
structure of $\gamma^{(2)}_{N,t}$ (which describes the
correlations among the particles) disappears when the weak limit
is taken, so that $\gamma_{\infty,t}^{(2)}$ lives on a length
scale of order one. Therefore, in (\ref{eq:GPwrong}) we get the
wrong coupling constant because we erroneously disregarded the
effect of the correlations present in $\gamma_{N,t}^{(2)}$. It is
hence clear that in order to derive the Gross-Pitaevskii equation
(\ref{eq:GPE2}) with the correct coupling constant $8\pi a_0$, we
need to take into account the short scale structure of
$\gamma^{(2)}_{N,t}$. To this end we begin by studying the ground
state of the system.

\medskip

A good approximation for the ground state wave function of the $N$
boson system is given by
\begin{equation*}
W(\bx) = \prod_{i<j}^N f(N(x_i -x_j))
\end{equation*}
where $f(x)$ is defined by (\ref{eq:scatt}) (then $f(Nx)$ solves
the same equation (\ref{eq:scatt}) with $V$ replaced by $V_a$).
Since we assumed the potential to be compactly supported (let $R$
denote the radius of its support), we have $f(x)= 1 - a_0 /|x|$,
for $|x|
> R$, and thus $f(Nx) = 1 -a_0/ N|x| = 1 - a/|x|$, for $|x|>Ra$. A
similar ansatz for the ground state wave function was already used
by Dyson in \cite{Dy} to prove his upper bound on the ground state
energy. In order to describe states of the condensate, it seems
appropriate to consider wave functions of the form
\begin{equation*}
\psi_N (\bx) = W (\bx) \phi_N (\bx)
\end{equation*}
where $\phi_N (\bx)$ varies over distances of order one, and is
approximately factorized, that is $\phi_N (\bx) \simeq
\prod_{j=1}^N \phi (x_j)$. Assuming for the moment that this form
is preserved under the time-evolution we have
\begin{equation*}
\gamma_{N,t}^{(2)} (x_1 , x_2; x'_1, x'_2) \simeq f(N(x_1 - x_2))
f(N(x'_1 -x'_2)) \gamma_{N,t}^{(1)} (x_1 ; x'_1)
\gamma_{N,t}^{(1)} (x_2 ; x'_2)\, .
\end{equation*}
Thus, for finite $N$, $\gamma^{(2)}_{N,t}$ is not exactly
factorized and has a short scale structure given by the function
$f(Nx)$. When we consider the limit $N \to \infty$ of the second
term on the right hand side of (\ref{eq:gamma1}) we obtain
\begin{equation}\label{eq:delta}
\begin{split}
\lim_{N\to\infty} N \int \rd x_2 \, &V_a (x_1 -x_2)
\gamma_{N,t}^{(2)} (x_1 , x_2 ; x'_1 , x_2) \\ &\hspace{.5cm} =
\lim_{N\to \infty} N^3 \int \rd x_2 \, V (N (x_1 - x_2)) f(N(x_1
-x_2)) \gamma^{(1)}_{\infty,t} (x_1 ; x'_1)
\gamma^{(1)}_{\infty,t} (x_2; x_2) \\ &\hspace{.5cm}= 8\pi a_0
\gamma_{\infty,t}^{(1)} (x_1;x'_1) \gamma^{(1)}_{\infty,t}
(x_1;x_1)
\end{split}
\end{equation}
where we used Eq. (\ref{eq:eightpia0}) and the fact that
$\gamma^{(1)}_{N,t}$ lives on a scale of order one (and thus we
can replace it by $\gamma_{\infty,t}^{(1)}$ without worrying about
the correlations). This leads to the Gross-Pitaevskii equation for
$\gamma_{\infty,t}^{(1)}$,
\begin{equation*}
\begin{split}
i\partial_t \gamma_{\infty,t}^{(1)} (x_1; x'_1) = \; &
\left(-\Delta_{x_1} + \Delta_{x'_1}\right) \gamma_{\infty,t}^{(1)}
(x_1 ; x'_1) + 8 \pi a_0 \left( \gamma_{\infty,t}^{(1)} (x_1 ;
x_1) - \gamma_{\infty,t}^{(1)} (x'_1 ; x'_1)\right)
\gamma_{\infty,t}^{(1)} (x_1; x'_1)
\end{split}
\end{equation*}
which has the correct coupling constant in front of the non-linear
term.

Note that the factorization
\begin{equation*}
\gamma^{(2)}_{\infty,t} (x_1 , x_2 ; x'_1, x'_2) =
\gamma^{(1)}_{\infty,t} (x_1 ; x'_1) \gamma^{(1)}_{\infty,t} (x_2
; x'_2)
\end{equation*}
still holds true, because the short scale structure of
$\gamma^{(2)}_{N,t}$ vanishes when the weak limit $N \to \infty$
is taken. The short scale structure only shows up in the
Gross-Pitaevskii equation due to the singularity of the potential.

\medskip

In order to make this heuristic argument for the derivation of the
Gross-Pitaevskii equation rigorous, we are faced with two major
steps.
\begin{itemize}
\item[i)] In the first step we have to prove that the $k$-particle
marginal density in the limit $N \to \infty$ really has the short
scale structure we discussed above. That is we have to prove that,
for large $N$,
\begin{equation}\label{eq:short}
\gamma_{N,t}^{(k +1)} (\bx_{k+1} ; \bx'_{k+1}) \simeq \left(
\prod_{i<j}^{k+1} f(N (x_i -x_j)) f (N(x'_i - x'_j)) \right)
\gamma_{\infty,t}^{(k+1)} (\bx_{k+1}; \bx'_{k+1})
\end{equation}
where $\gamma^{(k+1)}_{\infty,t}$ is the limit of
$\gamma_{N,t}^{(k+1)}$ with respect to some suitable weak topology
(in the heuristic argument above we considered the case $k=1$,
here $k$ is an arbitrary fixed integer $k \geq 1$). Eq.
(\ref{eq:short}) would then imply that, as $N \to \infty$, the
last term on the r.h.s. of the BBGKY hierarchy (\ref{eq:BBGKY})
converges to
\begin{equation*}
\lim_{N \to \infty} N \int \rd x_{k+1} \, V_a (x_j -x_{k+1})
\gamma_{N,t}^{(k+1)} (\bx_{k+1} ; \bx'_{k+1})  = 8 \pi a_0
\gamma_{\infty,t}^{(k+1)} ( \bx_k, x_j; \bx'_k , x_j) \,.
\end{equation*}
Therefore, if we could also prove that the second term on the
r.h.s. of (\ref{eq:BBGKY}) vanishes in the limit $N \to \infty$
(as expected, because formally of the order $N^{-1}$), then it
would follow that the family $\gamma_{\infty,t}^{(k)}$ satisfies
the Gross-Pitaevskii hierarchy
\begin{equation}\label{eq:GPH}
\begin{split}
i \partial_t \gamma_{\infty,t}^{(k)}  &(\bx_k ; \bx'_k) =
\sum_{j=1}^k \left(-\Delta_j + \Delta'_j \right)
\gamma_{\infty,t}^{(k)} (\bx_k ; \bx'_k) \\ & + 8\pi a_0
\sum_{j=1}^k \int \rd x_{k+1} \, \left( \delta (x_{k+1} - x_j) -
\delta (x_{k+1} - x'_j) \right)  \gamma_{\infty,t}^{(k+1)} (\bx_k
, x_{k+1} ; \bx'_k , x_{k+1})
\end{split}
\end{equation}
for all $k \geq 1$. We already know that this infinite hierarchy
of equation has a solution. In fact the factorized family of
densities $\gamma^{(k)}_{\infty,t} (\bx_k ; \bx'_k) =
\prod_{j=1}^k \gamma^{(1)}_{\infty,t} (x_j; x'_j)$ is a solution
of (\ref{eq:GPH}) if and only if $\gamma^{(1)}_{\infty,t}$ solves
the Gross-Pitaevskii equation (\ref{eq:GPE2}). \item[ii)]
Secondly, we need to prove that the densities
$\gamma^{(k)}_{\infty,t}$ factorize, that is, that, for all $k
\geq 1$,
\begin{equation}\label{eq:factorize}
\gamma^{(k)}_{\infty,t} (\bx_k ; \bx'_k) = \prod_{j=1}^k
\gamma_{\infty,t}^{(1)} (x_j; x'_j)\, .
\end{equation}
Then, from (\ref{eq:GPH}) and (\ref{eq:factorize}), it would
follow that $\gamma^{(1)}_{\infty,t}$ is a solution of the
Gross-Pitaevskii equation (\ref{eq:GPE2}). Note that, since we
already know that (\ref{eq:GPH}) has a factorized solution, in
order to prove (\ref{eq:factorize}) it is enough to prove the
uniqueness of the solution of the infinite hierarchy
(\ref{eq:GPH}).
\end{itemize}
Unfortunately, due to the singularity of the $\delta$-function, we
are still unable to prove that (\ref{eq:GPH}) has a unique
solution and thus we cannot prove part ii) (the best result in
this direction is the proof of the uniqueness for the hierarchy
with a Coulomb singularity, see \cite{EY}). On the other hand we
can complete part~i) of our program, that is, we can prove that
any limit point $\{ \gamma^{(k)}_{\infty,t} \}_{k \geq 1}$ of the
family $\{ \gamma_{N,t}^{(k)}\}_{k=1}^N$ (with respect to an
appropriate weak topology), satisfies the infinite hierarchy
(\ref{eq:GPH}), provided we replace the original Hamiltonian $H$
with a slightly modified version $\wt H$, where we artificially
modify the interaction when a large number of particles come into
a region with diameter much smaller than the typical
inter-particle distance. Since $H$ agrees with $\wt H$, apart in
the very rare event (rare with respect to the expected typical
distribution of the particles) that many particles come very close
together, we don't expect this modification to change the
macroscopic dynamics of the system: but unfortunately we cannot
control this effect rigorously.

\medskip

Note that the Gross-Pitaevskii equation (\ref{eq:GPE1}) is a
nonlinear Hartree equation
\begin{equation}\label{eq:hartree}
i\partial_t u_t = -\Delta u_t + (V * |u_t|^2) u_t
\end{equation}
in the special case $V(x) = 8\pi a_0 \delta (x)$. In the
literature there are several works devoted to the derivation of
(\ref{eq:hartree}) from the $N$-body Schr\"odinger equation. The
first results were obtained by Hepp in \cite{Hepp}, for a smooth
potential $V(x)$, and by Spohn in \cite{Spohn}, for bounded
$V(x)$. Later, Ginibre and Velo extended these results to singular
potentials in \cite{GV}: their result is limited to coherent
initial states, for which the number of particles cannot be fixed.
In \cite{EY}, Erd\H os and Yau derived (\ref{eq:hartree}) for the
Coulomb potential $V(x) = \pm 1/|x|$. More recently, Adami,
Bardos, Golse and Teta obtained partial results for the potential
$V(x) =  \delta (x)$, which leads to the Gross-Pitaevskii
equation, in the case of one-dimensional systems; see \cite{ABGT}.

\section{The Main Result}
In this section we explain how we need to modify the Hamiltonian
and then we state our main theorem. In order to derive
(\ref{eq:GPH}) it is very important to find a good approximation
for the wave function of the ground state of the N boson system.
We need an approximation which reproduces the correct short scale
structure and, at the same time, does not become too singular (so
that error terms can be controlled). Our first guess
\begin{equation}\label{eq:gs}
W (\bx) = \prod_{i<j} f_a (x_i -x_j) = \prod_{i<j} f (N (x_i -
x_j))
\end{equation}
is unfortunately not good enough. First of all we need to cutoff
the correlations at large distances (we want $f_a (x)=1$ for $|x|
\gg a$). To this end we fix a length scale $\ell_1 \gg a$, and we
consider the Neumann problem on the ball $\{ x: |x| \leq \ell_1
\}$ (we will choose $\ell_1 = N^{-2/3 + \kappa}$ for a small
$\kappa >0$). We are interested in the solution of the ground
state problem
\begin{equation*}
(-\Delta + 1/2 V_a (x) ) (1 -w (x)) = e_{\ell_1} (1 -w(x))
\end{equation*}
on $\{ x: |x| \leq \ell_1 \}$, with the normalization condition
$w(x) = 0$ for $|x| = \ell_1$. Here $e_{\ell_1}$ is the lowest
possible eigenvalue. It is easy to check that, up to contributions
of lower order, $e_{\ell_1} \simeq 3 a / \ell_1^3$. We can extend
$w(x)$ to be identically zero, for $|x| \geq \ell_1$. Then
\begin{equation}\label{eq:2body}
\begin{split}
(-\Delta + 1/2 V_a (x)) (1 -w(x)) &= q(x) (1-w(x)), \quad
\text{with} \\  q(x) &\simeq \frac{3a}{\ell_1^3} \chi (|x| \leq
\ell_1)\, .
\end{split}
\end{equation}
For $a \ll |x| \ll \ell_1$, the function $1-w(x)$ still looks very
much like $1 - a/|x|$, but now it equals one, for $|x| \geq
\ell_1$. Replacing $f_a (x_i -x_j)$ by $1- w(x_i-x_j)$ in
(\ref{eq:gs}) is still not sufficient for our purposes. The
problem is that the wave function $\prod_{i <j} (1 - w(x_i -x_j))$
becomes very singular when a large number of particles come very
close together. We introduce another cutoff to avoid this problem.
We fix a new length scale $\ell \gg \ell_1 \gg a$, such that $\ell
\ll N^{-1/3}$ (that is $\ell$ is still much smaller than the
typical inter-particle distance: we will choose $\ell = N^{-2/5
-\kappa}$ for a small $\kappa >0$). Then, for fixed indices $i$
and $j$, and for an arbitrary fixed number $K \geq 1$, we cutoff
the correlation between particles $i$ and $j$ (that is we replace
$1- w(x_i -x_j)$ by one) whenever at least $K$ other particles
come inside a ball of radius $\ell$ around $i$ and $j$. In order
to keep our exposition as clear as possible we choose $K=1$, that
is we cutoff correlations if at least three particles come very
close together. But there is nothing special about $K=1$: what we
really need to avoid are correlations among a macroscopic number
of particles, all very close together. To implement our cutoff we
introduce, for fixed indices $i,j$, a function $F_{ij} (\bx)$ with
the property that
\begin{equation*}
\begin{split}
 F_{ij} (\bx) &\cong 1 \quad \text{if} \quad \left\{
\begin{array}{ll} |x_i -x_m| &\gg \ell \\ |x_j -x_m| &\gg \ell
\end{array} \right. \quad \text{for all } m \neq i,j
\\  F_{ij} (\bx) &\cong 0 \quad \text{otherwise.} \end{split}
\end{equation*}
Instead of using the wave function $\prod_{i<j}( 1 - w(x_i -x_j))$
we will approximate the ground state of the $N$ boson system by
\begin{equation}\label{eq:gs3}
W(\bx) = \prod_{i<j} \left( 1 - w(x_i -x_j) F_{ij} (\bx) \right)\,
\end{equation}
(the exact definition of $W(\bx)$ is a little bit more
complicated; see \cite{ESY}, Section 2.3). The introduction of the
cutoffs $F_{ij}$ in the wave function $W(\bx)$ forces us to modify
the Hamiltonian $H$. To understand how $H$ has to be modified, we
compute its action on $W (\bx)$. We have, using (\ref{eq:2body}),
\begin{equation*}
\begin{split}
W(\bx)^{-1} (HW)(\bx) = \; &\sum_{i,j} q(x_i -x_j) \\ &+
\sum_{i,j}
\left((1/2) V_a (x_i -x_j) - q(x_i -x_j)\right) (1 - F_{ij} (\bx)) \\
&+ \text{lower order contributions.}
\end{split}
\end{equation*}
The ``lower order contributions'' are terms containing derivatives
of $F_{ij}$: they need some control, but they are not very
dangerous for our analysis. On the other hand, the second term on
the r.h.s. of the last equation, whose presence is due to the
introduction of the cutoffs $F_{ij}$, still contains the potential
$V_a$ and unfortunately we cannot control it with our techniques.
Therefore, we artificially remove it, defining a new Hamiltonian
$\wt H$, by
\begin{equation*} \wt H = H - \sum_{i,j} \left((1/2) V_a (x_i -x_j)
- q(x_i -x_j)\right) (1 - F_{ij} (\bx)) \, .
\end{equation*}
Note that the new Hamiltonian $\wt H$ equals the physical
Hamiltonian $H$ unless three or more particles come at distances
less than $\ell \ll N^{-1/3}$. This is a rare event, and thus we
don't expect the modification of the Hamiltonian $H$ to change in
a macroscopic relevant way the dynamics of the system.

\medskip

Before stating our main theorem, we still have to specify the
topology we use in taking the limit $N \to \infty$ of the marginal
distributions $\gamma^{(k)}_{N,t}$. It is easy to check that, for
every $k \geq 1$, $\gamma^{(k)}_{N,t} (\bx_k; \bx'_k) \in L^2
(\Lambda^k \times \Lambda^k)$. This motivates the following
definition. For \( \Gamma = \{ \gamma^{(k)} \}_{k \geq 1} \in
\bigoplus_{k \geq 1} L^2 (\Lambda^k \times \Lambda^k) \), and for
a fixed $\nu >1$, we define the two norms
\begin{equation}\label{eq:norms}
\begin{split}
\| \Gamma \|_- &:= \sum_{k \geq 0} \nu^{-k} \| \gamma^{(k)} \|_2
\quad  \quad \text{and } \quad \quad \| \Gamma \|_+ := \sup_{k
\geq 1} \nu^k \| \gamma^{(k)} \|_2
\end{split}
\end{equation}
where $\|. \|_2$  denotes the $L^2$-norm on $\Lambda^k \times
\Lambda^k$. We have to introduce the parameter $\nu >1$ to make
sure that, for $\Gamma_{N,t} = \{ \gamma^{(k)}_{N,t} \}_{k=1}^N$,
the norm $\| \Gamma_{N,t} \|_-$ is finite (choosing $\nu$ large
enough, we find $\| \Gamma_{N,t} \|_- \leq 1$, uniformly in $N$
and $t$). We also define the Banach spaces \[ \cH_- := \{ \Gamma =
\{ \gamma^{(k)} \}_{k \geq 0} \in \bigoplus_{k \geq 1} L^2
(\Lambda^k \times \Lambda^k) : \| \Gamma \|_- < \infty \}\] and
\[\cH_+ := \{ \Gamma = \{ \gamma^{(k)} \}_{k \geq 0} \in \bigoplus_{k
\geq 1} L^2 (\Lambda^k \times \Lambda^k) : \lim_{ k \to \infty}
\nu^k \| \gamma^{(k)} \|_2 =0 \} \, .\] Then we have $(\cH_- , \|
. \|_- ) = (\cH_+ , \| . \|_+ )^* $. This induces a weak* topology
on $\cH_-$, with respect to which the unit ball $\cB_-$ of $\cH_-$
is compact (Banach-Alaouglu Theorem). Since the space $\cH_+$ is
separable, the weak* topology on the unit ball $\cB_-$ is
metrizable: we can find a metric $\rho$ on $\cH_-$, such that a
sequence $\Gamma_n \in \cB_-$ converges with respect to the weak*
topology if and only if it converges with respect to the metric
$\rho$. For a fixed time $T$, we will also consider the space
$C([0,T], \cB_-)$ of functions of $t \in [0,T]$, with values in
the unit ball $\cB_- \subset \cH_-$, which are continuous with
respect to the metric $\rho$ (or equivalently with respect to the
weak* topology of $\cH_-$). We equip $C([0,T], \cB_-)$ with the
metric \[ \wt \rho (\Gamma_1 (t), \Gamma_2 (t)) = \sup_{t \in
[0,T]} \rho (\Gamma_1 (t), \wt \Gamma_2 (t))\, . \] In the
following we will consider the families $\Gamma_{N,t} = \{
\gamma^{(k)}_{N,t} \}_{k=1}^N$ as elements of $C([0,T], \cB_-)$,
and we will study their convergence and their limit points with
respect to the metric $\wt \rho$. We are now ready to state our
main theorem.

\begin{theorem}\label{thm:main}
Assume $a= a_0/N$, $\ell_1 = N^{-2/3 + \kappa}$, $\ell = N^{-2/5
-\kappa}$, for some sufficiently small $\kappa >0$. Assume
\[ ( \psi_{N,0}, \wt H^2 \psi_{N,0} ) \leq C N^2 \, ,\] where
$(.,.)$ denotes the inner product on $L^2 (\bR^{3N},\rd \bx)$. Let
$\psi_{N,t}$, for $t \in [0,T]$, be the solution of the
Schr\"odinger equation \begin{equation}\label{eq:schrmod}
i\partial_t \psi_{N,t} = \wt H \psi_{N,t} \end{equation} with
initial data $\psi_{N,0}$. Then, if $\alpha = (\| V\|_1 + \|
V\|_{\infty} )$ is small enough (of order one) and $\nu
>1$ is large enough (recall that $\nu$ enters the definition of
the norms (\ref{eq:norms})), we have:
\begin{itemize}
\item[i)] $\Gamma_{N,t} = \{ \gamma_{N,t}^{(k)} \}_{k=1}^N$ has at
least one (non-trivial) limit point \( \Gamma_{\infty,t} = \{
\gamma_{\infty,t}^{(k)} \}_{k \geq 1} \in C ([0,T], \cB_-)\) with
respect to the metric $\wt \rho$. \item[ii)] For any limit point
$\Gamma_{\infty,t} = \{ \gamma_{\infty,t}^{(k)} \}_{k\geq 1}$ and
for all $k \geq 1$, there exists a constant $C$ such that
\begin{equation}\label{eq:keyest}
\tr \, (1 -\Delta_i) (1-\Delta_j) \gamma^{(k)}_{\infty ,t} \leq C
\end{equation}
for all $i \neq j$, $t \in [0,T]$. \item[iii)] Any limit point
$\Gamma_{\infty,t}$ satisfies the infinite Gross-Pitaevskii
hierarchy (\ref{eq:GPH}) when tested against a regular function
$J^{(k)} (\bx_k ; \bx'_k)$:
\begin{equation}\label{eq:GPhier}
\begin{split}
\langle J^{(k)} , \gamma^{(k)}_{\infty,t} \rangle =  \; &\langle
J^{(k)} , \gamma^{(k)}_{\infty,0} \rangle -i \sum_{j=1}^k \int_0^t
\rd s \langle J^{(k)} , (-\Delta_j + \Delta_j')
\gamma^{(k)}_{\infty,s} \rangle \\ &- 8i \pi a_0 \sum_{j=1}^k
\int_0^t \rd s \int \rd \bx_k \rd \bx'_k  \, J^{(k)} (\bx_k
;\bx'_k) \\ &\hspace{.5cm} \times \int \rd x_{k+1}  \left( \delta
(x_j -x_{k+1}) - \delta (x'_j -x_{k+1})\right)
\gamma^{(k+1)}_{\infty,s} (\bx_k , x_{k+1} ; \bx'_k ,x_{k+1}) \, .
\end{split}
\end{equation}
Here we use the notation $\langle J^{(k)}, \gamma^{(k)} \rangle =
\int \rd \bx_k \rd \bx'_k \, \overline{J^{(k)} (\bx_k; \bx'_k)} \,
\gamma^{(k)} (\bx_k; \bx'_k)$.
\end{itemize}
\end{theorem}
{\it Remarks.} \begin{itemize} \item[i)] The main assumption of
the theorem is the requirement that the expectation of $\wt H^2$
at $t=0$ is of order $N^2$. One can prove that this condition is
satisfied for $\psi_{N,0}(\bx) = W(\bx) \phi_N (\bx)$ and $\phi_N$
sufficiently smooth (see \cite{ESY}, Lemma D1). Physically, this
assumption guarantees that the initial wave function $\psi_{N,0}(
\bx)$ has the short-scale structure characteristic of $W(\bx)$
and, hence, that it describes, locally, a condensate. \item[ii)]
It is a priori not clear that the action of the delta-functions in
the Gross-Pitaevskii hierarchy (\ref{eq:GPhier}) is well defined.
This fact follows by the bound (\ref{eq:keyest}), which makes sure
that $\gamma^{(k)}_{\infty,t}$ is sufficiently smooth.\item[iii)]
We also need to assume that $\alpha = (\| V\|_{\infty} + \| V
\|_1)$ is small enough (but still of order one). This technical
assumption is needed in the proof of the energy estimate,
Proposition \ref{prop:enest}.
\end{itemize}

\section{Sketch of the Proof}

In this section we explain some of the main ideas used in the
proof of Theorem~\ref{thm:main}. Let $\psi_{N,t}$ be the solution
of the Schr\"odinger equation (\ref{eq:schrmod}) (with the
modified Hamiltonian $\wt H$). We can decompose $\psi_{N,t}$ as
\[ \psi_{N,t} (\bx) = W(\bx) \phi_{N,t} (\bx), \] where $W(\bx)$
is the approximation for the ground state wave function defined in
(\ref{eq:gs3}). This decomposition is always possible because
$W(\bx)$ is strictly positive.

\medskip

The main tool in the proof of Theorem \ref{thm:main} is an
estimate for the $L^2$-norm of the second derivatives of
$\phi_{N,t}$. This bound follows from the following \emph{energy
estimate}.
\begin{proposition}\label{prop:enest}
Assume $a=a_0 /N$, $\ell_1 = N^{-2/3+ \kappa}$ and $\ell = N^{-2/5
-\kappa}$ for $\kappa>0$ small enough, and suppose $\alpha =
(\|V\|_1 + \|V\|_{\infty})$ is sufficiently small. Then there
exists a constant $C > 0$ such that
\begin{equation*}
\begin{split}
\int \rd \bx \, |(\tilde{H} W \phi) (\bx)|^2  \geq \; &(C - o (1))
 \sum_{i,j=1}^N \int \rd \bx \, W^2 (\bx) | \nabla_i \nabla_j \phi (\bx)|^2  \\
&- o(1) \left( N \sum_{i=1}^N \int \rd \bx \, W^2 (\bx) |\nabla_i
\phi (\bx)|^2 + N^2 \int \rd \bx \, W^2 (\bx)|\phi (\bx)|^2
\right)\, ,
\end{split}
\end{equation*}
where $o(1) \to 0$ as $N\to \infty$.
\end{proposition}
{\it Remark. } The proof of this proposition is the main technical
difficulty in our analysis. It is in order to prove this
proposition that we need to introduce the cutoffs $F_{ij}$ in the
approximate ground state wave function $W(\bx)$, and that we need
to modify the Hamiltonian.

\medskip

Using the assumption that, at $t =0$, $(\psi_{N,0} , \wt H^2
\psi_{N,0}) \leq C N^2$, the conservation of the energy, and the
symmetry with respect to permutations, we immediately get the
following corollary.

\begin{corollary}\label{cor}
Suppose the assumptions of Proposition \ref{prop:enest} are
satisfied. Suppose moreover that the initial data $\psi_{N,0}$ is
symmetric with respect to permutations and \( ( \psi_{N,0} , \wt
H^2 \psi_{N,0} ) \leq C N^2 \, .\) Then there exists a constant
$C$ such that \begin{equation}\label{eq:cor} \int W^2 (\bx)
|\nabla_i \nabla_j \phi_{N,t} (\bx) |^2 \leq C \end{equation} for
all $i \neq j$, $t$ and all $N$ large enough.
\end{corollary}
{\it Remark.} The bound (\ref{eq:cor}) is not an estimate for the
derivatives of the whole wave function $\psi_{N,t}$. The
inequality \begin{equation}\label{eq:psi2} \int \rd \bx \,
|\nabla_i \nabla_j \psi (\bx)|^2 < C \end{equation} is wrong, if
$\psi$ satisfies $(\psi , \wt H^2 \psi) \leq C N^2$. In fact, in
order for $(\psi, \wt H^2 \psi)$ to be of order $N^2$, the wave
function $\psi (\bx)$ needs to have the short scale structure
characterizing $W(\bx)$. This makes (\ref{eq:psi2}) impossible to
hold true uniformly in $N$. Only after the singular part $W(\bx)$
has been factorized out, we can prove bounds like (\ref{eq:cor})
for the derivatives of the remainder. One of the consequences of
our energy estimate, and one of the possible interpretation of our
result, is that the separation between the singular part of the
wave function (living on the scale $1/N$) and its regular part is
preserved by the time evolution.

\medskip

Next we show how the important bound (\ref{eq:cor}) can be used to
prove Theorem \ref{thm:main}. According to the decomposition
$\psi_{N,t} (\bx) = W(\bx) \phi_{N,t} (\bx)$, we define, for
$k=1,\dots ,N$, the densities $U_{N,t}^{(k)} (\bx_k;\bx'_k)$, for
$k=1, \dots ,N$, to be, roughly, the $k$-particle marginal density
corresponding to the wave function $\phi_{N,t}$ (the exact
definition is a little bit more involved, see \cite{ESY}, Section
4). The estimate (\ref{eq:cor}) for the second derivatives of
$\phi_{N,t}$ translates into a bound for the densities
$U_{N,t}^{(k)}$:
\begin{equation}\label{eq:boundU} \tr \, (1 -\Delta_i)
(1-\Delta_j) U_{N,t}^{(k)} \leq C \end{equation} for all $i,j \leq
N$ with $i \neq j$, for all $t$ and for all $N$ large enough.

Moreover, we can show that, for $\nu >1$ large enough (recall that
the parameter $\nu$ enters the definition of the norms
(\ref{eq:norms})), the families $U_{N,t} = \{ U_{N,t}^{(k)}
\}_{k=1}^N$ define an equicontinuous sequence in the space
$C([0,T], \cB_-)$ (this follows from a careful analysis of the
BBGKY hierarchy associated to the Schr\"odinger equation
(\ref{eq:schrmod}); see \cite{ESY}, Sections 9.1 and 9.2 for more
details). Applying standard results (Arzela-Ascoli Theorem), it
follows that the sequence $U_{N,t}$ has at least one limit point,
denoted $U_{\infty,t} = \{ U_{\infty,t}^{(k)} \}_{k\geq 1}$, in
the space $C([0,T], \cB_-)$. The bound (\ref{eq:boundU}) can then
be passed to the limit $N\to \infty$, and we obtain
\[ \tr \, (1 - \Delta_i)(1- \Delta_j) U_{\infty,t}^{(k)} \leq C \]
for all $i \neq j$ and $t \in [0,T]$.

Next we go back to the family $\Gamma_{N,t} = \{
\gamma_{N,t}^{(k)} \}_{k=1}^N$. Clearly, the densities
$\gamma^{(k)}_{N,t}$ do not satisfy the estimate
(\ref{eq:boundU}). In fact, $\gamma_{N,t}^{(k)}$ still contains
the short scale structure of $W(\bx)$ (which, on the contrary, has
been factorized out from $U_{N,t}^{(k)}$), and thus cannot have
the smoothness required by (\ref{eq:boundU}).

It is nevertheless clear that the short scale structure of
$\Gamma_{N,t}$ disappears when we consider the limit $N \to
\infty$ (in the weak sense specified by Theorem~\ref{thm:main}).
In fact, one can prove the convergence of an appropriate
subsequence of $\Gamma_{N,t}$ to the limit point $U_{\infty,t}$ of
$U_{N,t}$. In other words one can show that limit points of
$\Gamma_{N,t}$, denoted by $\Gamma_{\infty,t}$, coincide with the
limit points of $U_{N,t}$. Therefore, even though $\Gamma_{N,t}$,
for finite $N$, does not satisfies the bound (\ref{eq:boundU}),
its limit points $\Gamma_{\infty,t} = \{ \gamma^{(k)}_{\infty,t}
\}_{k \geq 1}$ do. For every $k\geq 1$ we have
\begin{equation}\label{eq:boundGamma}
\tr \, (1 -\Delta_i) (1-\Delta_j) \gamma_{\infty,t}^{(k)} \leq C
\end{equation}
for all $i \neq j$ and $t\in [0,T]$. This proves part i) and ii)
of Theorem \ref{thm:main} (the non-triviality of the limit follows
by showing that $\tr \, \gamma^{(1)}_{\infty,t} =1$). The bound
(\ref{eq:boundGamma}) can then be used to prove part iii) of
Theorem \ref{thm:main}, that is to prove that the family
$\Gamma_{\infty,t}$ satisfies the infinite Gross-Pitaevskii
hierarchy (\ref{eq:GPhier}). In fact, having control over the
derivatives of $\gamma^{(k)}_{\infty,t}$ allows us to prove the
convergence of the potential to a delta-function (that is, it
allows us to make (\ref{eq:delta}) rigorous). To this end we use
the following lemma (see \cite{ESY}, Section 8).
\begin{lemma}
Suppose $\delta_{\beta} (x) = \beta^{-3} h (x/\beta)$, for some
regular function $h$, with $\int h(x) =1$. Then, for any $1\leq j
\leq k$, and for any regular function $J(\bx_k; \bx'_k)$, we have
\begin{multline*}
\Big| \int \rd {\bf x}_k \rd {\bf x}_k' \rd x_{k+1} \, J ({\bf
x}_k ; {\bf x}_k') (\delta_{\beta} (x_j -x_{k+1}) - \delta (x_j
-x_{k+1}) ) \gamma^{(k+1)} (\bx_k,x_{k+1};
\bx'_k , x_{k+1}) \Big| \\
\leq C_J  \, \sqrt \beta
 \, \tr \, (1 -\Delta_j) (1 -\Delta_{k+1}) \gamma^{(k+1)} \; .
\end{multline*}
\end{lemma}
Part iii) of Theorem \ref{thm:main} can then be proven combining
this lemma with the estimates (\ref{eq:boundU}) and
(\ref{eq:boundGamma}) (see \cite{ESY}, Section 9.4, for more
details).


\thebibliography{hh}

\bibitem{ABGT} R. Adami, C. Bardos, F. Golse and
 A. Teta: {\sl Towards a rigorous derivation of the
cubic nonlinear Schr\"odinger equation in dimension one.}
Asymptot. Anal. (2) {\bf 40} (2004), 93--108.

\bibitem{Dy} F.J. Dyson: {\sl Ground-state energy of a hard-sphere
gas. \/} Phys. Rev. (1) {\bf 106} (1957), 20--26.

\bibitem{EESY} A. Elgart, L. Erd{\H{o}}s, B. Schlein, and H.-T. Yau:
{\sl Gross-Pitaevskii equation as the mean field limit of weakly
coupled bosons. \/} Preprint math-ph/0410038. To appear in Arch.
Rat. Mech. Anal.

\bibitem{ESY} L. Erd{\H{o}}s, B. Schlein, and H.-T. Yau: {\sl
Derivation of the {G}ross--{P}itaevskii hierarchy for the dynamics
of a Bose-Einstein condensate. \/} Preprint math-ph/0410005.

\bibitem{EY} L. Erd{\H{o}}s and H.-T. Yau: {\sl Derivation
of the nonlinear {S}chr\"odinger equation from a many body
{C}oulomb system.} Adv. Theor. Math. Phys. (6) {\bf 5} (2001),
1169--1205.

\bibitem{GV} J. Ginibre and G. Velo: {\sl The classical
field limit of scattering theory for non-relativistic many-boson
systems. I and II.} Commun. Math. Phys. {\bf 66} (1979), 37--76
and {\bf 68} (1979), 45--68.

\bibitem{G1} E.P. Gross: {\sl Structure of a quantized vortex in boson systems.}
Nuovo Cimento {\bf 20} (1961), 454--466.

\bibitem{G2} E.P. Gross: {\sl Hydrodynamics of a superfluid condensate. \/}
J. Math. Phys. {\bf 4} (1963), 195--207.

\bibitem{Hepp} K. Hepp: {\sl The classical limit for quantum mechanical
correlation functions. \/} Commun. Math. Phys. {\bf 35} (1974),
265--277.

\bibitem{LS} E.H. Lieb and R. Seiringer: {\sl
Proof of {B}ose-{E}instein Condensation for Dilute Trapped Gases.
\/} Phys. Rev. Lett. {\bf 88} (2002), 170409-1-4.

\bibitem{LSSY2}  E.H. Lieb, R. Seiringer, J.P. Solovej, and J. Yngvason: {\sl
The Quantum-Mechanical Many-Body Problem: {B}ose Gas.\/} Preprint
math-ph/0405004.

\bibitem{LSY1} E.H. Lieb, R. Seiringer, J. Yngvason: {\sl Bosons in a Trap:
A Rigorous Derivation of the {G}ross-{P}itaevskii Energy
Functional.} Phys. Rev A {\bf 61} (2000), 043602.

\bibitem{LY1} E.H. Lieb and J. Yngvason: {\sl Ground State Energy of the low
density {B}ose Gas.} Phys. Rev. Lett. {\bf 80} (1998), 2504--2507.

\bibitem{P} L.P. Pitaevskii: {\sl Vortex lines in an imperfect {B}ose
gas. \/} Sov. Phys. JETP {\bf 13} (1961), 451--454.

\bibitem{Spohn} H. Spohn:
{\sl Kinetic Equations from Hamiltonian Dynamics.}
    Rev. Mod. Phys. {\bf 52} no. 3 (1980), 569--615.
\end{document}